\documentclass[aps,prd,preprintnumbers,superscriptaddress,nofootinbib,amsmath,amssymb,showpacs,preprint]{revtex4}
\usepackage{graphicx}
\usepackage{dcolumn}
\usepackage{bm}
\usepackage{amsmath}
\usepackage{color}
\usepackage{ulem}

\input{colordvi.tex}

\newcommand{\gsim}{~\mbox{\raisebox{-1.0ex}{$\stackrel{\textstyle >}
{\textstyle \sim}$ }}}
\newcommand{\lsim}{~\mbox{\raisebox{-1.0ex}{$\stackrel{\textstyle <}
{\textstyle \sim}$ }}}

\newcommand{\beq}{\begin{equation}}
\newcommand{\eeq}{\end{equation}}
\newcommand{\beqa}{\begin{eqnarray}}
\newcommand{\eeqa}{\end{eqnarray}}

\begin{document}

\begin{flushright}
CTPU-PTC-18-06
\end{flushright}

\title{Magnetic Field Transfer From A Hidden Sector}

\author{Kohei Kamada}
\email[Email: ]{kkamada"at"ibs.re.kr}
\affiliation{Center for Theoretical Physics of the Universe, Institute for Basic Science (IBS), 
   Daejeon, 34126, Korea}
\affiliation{School of Earth and Space Exploration, Arizona State University, Tempe, AZ 85287, USA}

\author{Yuhsin Tsai}
\email[Email: ]{yhtsai"at"umd.edu}
\affiliation{Maryland Center for Fundamental Physics, Department of Physics,
University of Maryland, College Park, MD 20742, USA}

\author{Tanmay Vachaspati}
\email[Email: ]{tvachasp"at"asu.edu}
\affiliation{Physics Department, Arizona State University, Tempe, AZ 85287, USA}


\begin{abstract}
Primordial magnetic fields in the dark sector can be transferred to magnetic 
fields in the visible sector due to a gauge kinetic mixing term. 
We show that the transfer occurs when the evolution of magnetic fields is dominated by
dissipation due to finite electric conductivity, and does not occur at later
times if the magnetic fields evolve according to magnetohydrodynamics scaling laws. 
The efficiency of the transfer is suppressed by not only the gauge kinetic mixing coupling 
but also the ratio between the large electric conductivity and the typical momentum of the 
magnetic fields. 
We find that the transfer gives nonzero visible magnetic fields today. 
However, without possible dynamo amplifications, the field transfer is not efficient enough to obtain the intergalactic 
magnetic fields suggested by the gamma-ray observations, although there are plenty of possibilities for efficient dark 
magnetogenesis, which are experimentally unconstrained.
\end{abstract}
\maketitle

\section{Introduction}

Primordial magnetic fields have been of interest for many years since they may
explain the observed galaxy and galaxy cluster magnetic fields 
through the dynamo mechanism during structure formation~\cite{Widrow:2002ud}.  
Moreover, the presence of intergalactic magnetic fields is also indicated by the recent
observations of TeV blazars~\cite{Neronov:1900zz,Tavecchio:2010mk,Ando:2010rb,Dolag:2010ni,
Essey:2010nd,Taylor:2011bn,Takahashi:2013lba,Finke:2015ona}, which provide a lower bound on the magnetic field
strength, $B\gtrsim 10^{-19}$ G at Mpc coherence scales and 
$B \gtrsim 10^{-16}$ G $\times (\lambda/{\rm pc})^{-1/2}$ at smaller length scales~\cite{Finke:2015ona}. 
However, it is difficult to come up with astrophysical origins of these magnetic fields 
in the cosmic voids, and the challenge motivates the consideration of these intergalactic magnetic fields as remnants 
from the very early universe~\cite{Vachaspati:2016xji}. Such magnetic fields can even be related to the matter-antimatter asymmetry of the 
Universe~\cite{Giovannini:1997gp,Giovannini:1997eg,Bamba:2006km,Fujita:2016igl,Kamada:2016eeb,Kamada:2016cnb} 
or the production of dark matter density~\cite{Kamada:2017cpk}.

There have been many proposals for primordial magnetogenesis, 
such as the inflationary magnetogenesis~\cite{Turner:1987bw,Ratra:1991bn,Garretson:1992vt,Anber:2006xt}, 
productions from the first order phase transition of the electroweak symmetry~\cite{Vachaspati:1991nm,Baym:1995fk,Grasso:1997nx} 
or QCD~\cite{Quashnock:1988vs,Cheng:1994yr,Sigl:1996dm}, 
or productions through the chiral instability~\cite{Joyce:1997uy,Tashiro:2012mf,Brandenburg:2017rcb}. It remains
to be seen if these proposals can match magnetic field spectra as indicated by the blazar observations
(see, {\it e.g.}, Refs.~\cite{Durrer:2013pga,Grasso:2000wj,Subramanian:2015lua} for reviews). 
In fact, it has been noticed that almost all the existing magnetogenesis proposals have problems in 
addressing the blazar issue.
For example, inflationary magnetogenesis models are 
strongly constrained by observations of cosmic microwave background (CMB),
which make it difficult to generate the required magnetic fields in these scenarios 
(see, however, Ref.~\cite{Fujita:2016qab}).
The electroweak and QCD phase transitions are known to be 
crossovers within the Standard Model (SM)~\cite{Aoki:2006we,Kajantie:1996mn}, 
and it is not clear if magnetic fields could be generated by the SM phase transitions.

However, once we consider particle physics beyond the SM, 
there are much more possibilities of magnetogenesis from the existence of additional U$(1)$ symmetries, which can be preserved at an earlier time universe and therefore suffer from  weaker experimental constraints. The additional U$(1)$ can be a gauged U$(1)_{B-L}$, other U$(1)$ symmetries arising in grand unified theories~\cite{Jeannerot:2003qv}, or simply a dark U$(1)$ field that couples weakly to the visible sector.
We can imagine a ``dark magnetogenesis" mechanism in a hidden sector, 
for example, from a much stronger first order cosmological phase transition than the SM symmetry breaking.
The dark symmetry breaking into a dark U$(1)$ gauge symmetry can produce a strong dark magnetic field, which 
is mildly constrained if the process happens after inflation but before Big-bang nucleosynthesis (BBN), and the dark photon later obtains a mass from the dark U$(1)$ breaking and decays into SM particles before BBN. Although dark magnetic monopoles may also be generated during the phase transition, we will not consider the case by assuming that the larger gauge symmetry group has a nontrivial first homotopy. 
If the dark magnetic fields are transferred to the SM magnetic fields after their production, 
they may provide seeds for the galaxy and galaxy cluster magnetic fields and explain 
the TeV blazar observation. In this article we examine the evolution of dark magnetic fields 
and how a transfer from dark to electromagnetic (or hypermagnetic in the SM) magnetic fields 
can occur. A related idea in which background dark photon generated through an oscillating axion-like particle gets converted into visible magnetic field is also discussed recently in Ref.~\cite{Choi:2018dqr}.

Regardless of the details of the model, the dark U$(1)_D$ gauge field $D_\mu$ and the visible U$(1)_Y$ gauge 
field $Y_\mu$ will interact via a {\it gauge kinetic mixing} term, $-\epsilon D_{\mu\nu} Y^{\mu\nu}$, 
with $\epsilon$ being the gauge kinetic mixing parameter. 
Such a gauge kinetic mixing can be removed by field redefinition but generally only
when there are no couplings to matter. Once we introduce couplings to matter fields, the visible and dark gauge fields 
as well as the gauge kinetic mixing 
are uniquely defined. 
Here we define the gauge fields so that  the SM matter fields are not charged under 
the dark U(1) symmetry in the basis with the nontrivial gauge kinetic mixing. 

To study the cosmological evolution of dark and visible magnetic fields, we must account for
the plasma in which these magnetic fields are embedded. Therefore, instead of solving the classical field theory equations, we study magnetohydrodynamic (MHD) equations that have been extended to include the dark sector fields. 
With some simplifying assumptions, notably ignoring 
turbulence, we find that dark magnetic fields 
are transferred to the visible sector at early times. The transfer efficiency is suppressed by a factor
of $\epsilon k_c^2 \Delta t_{\rm s} /\sigma_Y$ with 
$k_c$ being the typical momentum of the dark magnetic fields, $\Delta t_{\rm s}$ being 
the duration of the transfer, and $\sigma_Y$ being the (hyper)electric 
conductivity. At late times, once the magnetic fields evolve
according to scaling laws indicated by MHD simulations, no further transfer occurs if there is no dynamo amplification. 
Unless further amplification of the magnetic field occurs, {\it e.g.} at the time
of dark U(1) symmetry breaking, the suppression factor implies that the visible fields
are too weak to explain the blazar observations.

The paper is organized as follows. In Sec.~\ref{sec2} we describe the model
and derive the evolution equations of visible and dark magnetic fields. 
In Sec.~\ref{sec3} we examine how the transfer from dark to visible magnetic fields occurs. 
In Sec.~\ref{sec4} we adopt the formalism developed in the previous sections and evaluate the present properties of the intergalactic magnetic fields. We summarize our findings in Sec.~\ref{sec5}.

\section{Model and evolution equations \label{sec2}}

We focus on the case where there are no light matter fields in the plasma that are 
simultaneously charged under both the visible and dark U(1) symmetries in the basis 
with gauge kinetic mixing. 
(The case with particles in the plasma that are charged under both U(1) symmetries
is discussed in Appendix~\ref{app1}.)
The Lagrangian is now written as
\begin{equation}
{\cal L}= -\frac{1}{4} Y_{\mu\nu} Y^{\mu\nu} -\frac{1}{4} D_{\mu\nu} D^{\mu\nu} -\frac{\epsilon}{2}Y_{\mu\nu} D^{\mu\nu} - J_Y^\mu Y_\mu - J_D^\mu D_\mu. \label{lag}
\end{equation}
Here $Y_\mu$ is the SM hypercharge gauge field, $D_\mu$ is the dark U(1) gauge field, and 
$Y_{\mu\nu}$ and $D_{\mu\nu}$ are their field strengths. 
$\epsilon \ll 1$ is the gauge kinetic mixing parameter, which can come from a loop-induced process with heavy mediators 
connecting the two sectors. $J_\mu^Y$ and $J_\mu^D$ are the visible and dark U(1) current carried by matter fields 
with the associated U(1) charges. 
We assume both U(1) symmetries remain unbroken throughout the $B$-field transferring process. 
Depending on the mass of the dark photon, there are constraints on $\epsilon$ \cite{Alexander:2016aln}.
However, for a high dark U(1) breaking scale, much higher than the electroweak scale, 
there are no strong bounds on $\epsilon$ and hence $\epsilon \sim {\cal O}(0.1)$ 
is allowed.
Since we focus on the dynamics at a scale higher than the electroweak scale, 
the visible magnetic fields are identified as the SM hyper U(1) magnetic fields. 
Hyper magnetic fields are subsequently transformed into (electro)magnetic fields at the electroweak
phase transition~\cite{Kamada:2016cnb}.

Here we consider the case where the Universe is filled with thermal fluids, in which both U(1) charged particles are thermalized. 
In such an environment, the evolution of magnetic fields with a spatial scale larger than the intrinsic scale of the fluids can be described by the MHD equations~\cite{Biskamp}, which consist of the Navier-Stokes equations and Maxwell's equations. 
We modify the Maxwell's equations to include both gauge fields with kinetic mixing. 

The focus of this work is the transition between dark and SM magnetic fields inside thermal fluids. 
Instead of giving a specific $B$-field generation model in the dark sector, we simply assume the existence 
of a dark magnetic field, $B_D(t_i)$, from the initial conformal time $t_i$. The dark $B$-field can come from various field 
generation models but with a larger parameter space for phase 
transitions or chiral instability that are not directly constrained by SM physics.

The Lagrangian Eq.~\eqref{lag} leads to the equations of motion for the gauge fields,
\begin{equation}
\partial_\mu Y^{\mu\nu}+\epsilon \,\partial_\mu D^{\mu\nu} = J_Y^\nu, \quad \partial_\mu D^{\mu\nu}+\epsilon\, \partial_\mu Y^{\mu\nu} = J_D^\nu. 
\end{equation}
In terms of the electric and magnetic fields, we obtain modified Amp\'ere's laws
as
\begin{equation}
{\bm \nabla}\times ({\bm B}_Y+\epsilon \,{\bm B}_D) = {\bm J}_Y, \quad {\bm \nabla}\times ({\bm B}_D+\epsilon \,{\bm B}_Y) = {\bm J}_D,   \label{max1}
\end{equation}
where ${\bm B}_Y$ and ${\bm B}_D$ are the magnetic fields for the visible and dark gauge 
fields. We work in the conformal frame so that the effects of the cosmic expansion in the 
Friedmann Universe do not appear explicitly. The time should be understood as the conformal time 
and the electric conductivities should be rescaled by the scale factor $a$,  $\sigma_a = a\,\sigma_{a, {\rm phys}}$.
The physical electric and magnetic fields,  ${\bm E}_{\rm phys}$ and ${\bm B}_{\rm phys}$, are obtained 
by ${\bm E}_{\rm phys} = a^{-2} {\bm E}$ and ${\bm B}_{\rm phys} = a^{-2} {\bm B}$. We have adopted the nonrelativistic MHD approximation 
and neglected the displacement currents ${\dot {\bm E}}_Y$ and ${\dot {\bm E}}_D$,
where ${\bm E}_Y$ and ${\bm E}_D$ are the visible and dark electric fields,
since they are suppressed by factors of the fluid velocity $v\ll1$ compared to the total 
currents~\cite{Choudhuri}. 
Faraday's laws as well as Gauss's laws for magnetism take the standard form,
\begin{equation}
{\bm \nabla}\times {\bm E}_Y=-{\dot {\bm B}}_Y \quad {\bm \nabla}\times {\bm E}_D=-{\dot {\bm B}}_D, \quad {\bm \nabla} \cdot {\bm B}_Y=0, \quad {\bm \nabla} \cdot {\bm B}_D=0, \label{max2}
\end{equation}
since they are derived from the definition of the field strength tensor. 

Assuming the chiral magnetic 
current~\cite{Vilenkin:1980fu} as well as the chiral vortical current~\cite{Vilenkin:1979ui}  
are negligibly small, the currents obey Ohm's law,
\begin{equation}
{\bm J}_a = \sigma_{ab} \left({\bm E}_b + {\bm v} \times {\bm B}_b \right), \quad a,b = Y,\,D, 
\end{equation}
where ${\bm v}$ is the local velocity of both the SM and dark fluids,  
and $\sigma_{ab}$ is the electric conductivity tensor. 
To justify the treatment of the medium as a single fluid, we note that
$t$-channel scattering between dark and SM particles, assuming similar masses, 
keeps the visible and dark fluids in thermal equilibrium as long as the
scattering rate $\Gamma \sim N_{\rm scat} \epsilon^2 \alpha_{Y}^2 T$, is
larger than the Hubble expansion rate $H\sim T^2/ m_{\rm Pl}$ 
where $N_{\rm scat}$, $\alpha_{Y}$ and $m_{\rm Pl}=1.22\times 10^{19}$ GeV are the 
number of particles that are involved in the scattering, the hyper fine structure constant and 
the Planck mass, respectively. Thus in order for the single fluid approximation to be justified, the temperature of the fluid
must be smaller than 
\begin{equation}
T\lesssim 10^{14} {\rm GeV} \left(\frac{N_{\rm scat}}{100}\right) \left(\frac{\epsilon}{0.1}\right)^2.  \label{sfapp}
\end{equation}
At much lower temperatures, 
either the dark U$(1)$ breaking or recombinations in the two sectors makes the system 
depart from thermal equilibrium.

In the high temperature phase, the conductivity tensor is evaluated by the Kubo formula as
\begin{equation}
\sigma_{ab} = - \lim_{\omega \rightarrow 0} \lim_{k\rightarrow 0} \frac{1}{\omega} {\rm Im} \langle J_a J_b \rangle_{\rm irr}
\end{equation}
with the bracket being the one-boson irreducible correlation function~\cite{Baym:1997gq,Arnold:2000dr}. 
In our setup, since there are no fields that carry both the visible and dark U(1) charges, 
the off-diagonal components of the electric conductivity tensor vanishes at tree level 
and is suppressed by the kinetic mixing $\epsilon$ at higher order.
Neglecting the off-diagonal components
(see Appendix A for details), 
we write the visible and dark electric currents in terms of the visible and dark electric and magnetic 
fields as
\begin{equation}
{\bm J}_Y= \sigma_Y ({\bm E}_Y+{\bm v}\times {\bm B}_Y), \quad {\bm J}_D= \sigma_D ({\bm E}_D+{\bm v}\times {\bm B}_D). \label{ohm1}
\end{equation}
The visible and dark electric conductivities 
($\sigma_Y \equiv \sigma_{YY}$ and $\sigma_D \equiv \sigma_{DD}$) 
are evaluated as~\cite{Baym:1997gq,Arnold:2000dr}
\begin{equation}\label{eq:conduc}
\sigma_Y \sim \sigma_D \sim a \,C\,\frac{T^3\,g^2}{T\,(g^4\,T\,\ln g^{-1})}\sim 10^2\left(\frac{C}{10}\right)\left(\frac{e}{g}\right)^{2}\,a \,T,
\end{equation} 
with $g$ being the gauge coupling of the dominant thermal fluid particles, and $e$ is the SM electric charge. The equation can be qualitatively understood as arising from the classical Drude model, $\sigma\sim n\,g^2\,\tau/m$, with number density $n\sim T^3$, typical energy scale $m\sim T$, and the characteristic time scale for large angle scattering $\tau\sim(g^4\,T\,\ln g^{-1})^{-1}$ in thermal bath. The coefficient $C$ depends on the number of charged particle species, 
and in the SM ranges from 15, when only the electron is included, to 12, when all charged fermions besides top are included~\cite{Arnold:2000dr}. 
Here the scale factor $a$ is included since we define the electric conductivities in the conformal frame. As a result,  the electric conductivities are invariant under the cosmic expansion in the limit we neglect the change of the number of relativistic particles.

We can eliminate the electric fields from Ohm's and Amp\'ere's laws,
so that the evolution equations for the magnetic fields read
\begin{align}
{\dot {\bm B}_Y} &= \frac{1}{\sigma_Y} {\bm \nabla}^2 {\bm B}_Y + \frac{\epsilon}{\sigma_Y} {\bm \nabla}^2 {\bm B}_D+{\bm \nabla}\times ({\bm v} \times {\bm B}_Y) \label{ees1}, \\
{\dot {\bm B}_D} &=  \frac{1}{\sigma_D} {\bm \nabla}^2 {\bm B}_D + \frac{\epsilon}{\sigma_D} {\bm \nabla}^2 {\bm B}_Y+{\bm \nabla}\times ({\bm v} \times {\bm B}_D).  \label{ees2}
\end{align}
By redefining 
\begin{align}
{\hat {\bm B}}_Y &\equiv -\dfrac{\epsilon}{\sqrt{(1-\alpha)^2+4 \alpha \epsilon^2}} {\bm B}_Y +\left( \dfrac{1}{2}-\dfrac{1-\alpha}{2\sqrt{(1-\alpha)^2+4 \alpha \epsilon^2}}\right) {\bm B}_D, \label{bas1}\\
 {\hat {\bm B}}_D &\equiv \dfrac{\epsilon}{\sqrt{(1-\alpha)^2+4 \alpha \epsilon^2}}{\bm B}_Y+\left( \dfrac{1}{2}+\dfrac{1-\alpha}{2\sqrt{(1-\alpha)^2+4 \alpha \epsilon^2}}\right) {\bm B}_D, \label{bas2}
\end{align}
with $\alpha \equiv \sigma_D/\sigma_Y$, 
the evolution equations for the magnetic fields are decoupled as
\begin{align}\label{eq:evoeq1}
{\dot {\hat {\bm B}}_Y} &= \dfrac{1+\alpha-\sqrt{(1-\alpha)^2+4 \alpha \epsilon^2}}{2 \alpha}  \frac{{\bm \nabla}^2}{\sigma_Y} {\hat {\bm B}}_Y +{\bm \nabla}\times ({\bm v} \times {\hat {\bm B}}_Y ), \\
\label{eq:evoeq2}
{\dot {\hat {\bm B}}_D} &=  \dfrac{1+\alpha+\sqrt{(1-\alpha)^2+4 \alpha \epsilon^2}}{2 \alpha}    \frac{{\bm \nabla}^2}{\sigma_Y} {\hat {\bm B}}_D +{\bm \nabla}\times ({\bm v} \times {\hat {\bm B}}_D ).
\end{align}
We can see that the nonvanishing gauge kinetic mixing and/or $\alpha\neq 1$ generate a difference in the 
effective electric conductivities and therefore a different time evolution of the two magnetic fields.
This will be the source for the transfer from the dark to visible magnetic fields as we show
in the next section by solving the evolution equations.
Note that the field redefinition (Eqs.~\eqref{bas1} and \eqref{bas2}) makes sense 
only when $\epsilon \neq 0$ and hence nonvanishing $\epsilon$ is essential for the 
magnetic field transfer.

\section{Transfer of magnetic fields \label{sec3}} 

We explore the transfer of dark magnetic fields to visible magnetic fields in two steps. 
First, at early times, soon after the dark $B_D$ is generated, we assume that the fluid velocity is negligible.
At such early times, energy in the $B$-fields has not been transferred to kinetic flows and this assumption
is justified. 
Eventually the velocity fields are emerged through the Lorentz force and the 
eddy turnover scale catches up with the coherence scale of the $B$-fields. 
At that time we can no longer ignore the fluid velocity. 
In this second stage, however, we can use the 
scaling laws derived using numerical MHD simulations~\cite{Banerjee:2004df,Kahniashvili:2012uj}.

\subsection{First stage: $v \approx 0$}

Setting $v \to 0$ in Eqs.~(\ref{eq:evoeq1}) and (\ref{eq:evoeq2}), the equations linearize.
Then it is convenient to go to Fourier space, 
\begin{equation}
{\bm B}(t, {\bm x}) = \int \frac{d^3 k}{(2\pi)^3} \sum_{s\,= \pm} B^s (k,t) {\bm Q}^s({\bm k}) \exp[i{\bm k} \cdot {\bm x}],  
\end{equation}
with ${\bm Q}^{\pm}({\bm k})$ being the circular polarization vectors. 
The mode functions $B^s (k,t)$ then obey,
\begin{align}
{\dot {\hat B}^s_Y} (k,t) &= -  \dfrac{1+\alpha-\sqrt{(1-\alpha)^2+4 \alpha \epsilon^2}}{2 \alpha} \frac{k^2}{\sigma_Y} {\hat B^s}_Y(k,t), \\
{\dot {\hat B}^s_D} (k,t) &= -  \dfrac{1+\alpha+\sqrt{(1-\alpha)^2+4 \alpha \epsilon^2}}{2 \alpha} \frac{k^2}{\sigma_Y} {\hat B^s}_D(k,t),
\end{align}
with the solutions,
\begin{align}
{\hat B}^s_Y (k,t) &=\exp\left[ -  \dfrac{1+\alpha-\sqrt{(1-\alpha)^2+4 \alpha \epsilon^2}}{2 \alpha} \frac{k^2}{\sigma_Y}(t-t_i)\right] {\hat B}^s_Y (k,t_i), \label{sol1}\\
{\hat B}^s_D (k,t) &=\exp\left[ -  \dfrac{1+\alpha+\sqrt{(1-\alpha)^2+4 \alpha \epsilon^2}}{2 \alpha} \frac{k^2}{\sigma_Y}(t-t_i)\right] {\hat B}^s_D (k,t_i), \label{sol2}
\end{align}
where $t$ is the conformal time, and $t_i$ is the time of $B_D$ generation. The exponential decay corresponds to dissipation of the $B$-fields caused by the finite conductivity. ${\hat B}^s_Y$ and ${\hat B}^s_D$ decay with different rates due to the different effective conductivities. Since we are interested in having both the initial dark $B$-field generation and the later time dark U$(1)$ breaking in the un-hatted basis ($B_{Y,D}$), we write the solutions in the original basis as in Eq.~(\ref{lag}) and take $B_Y=0$ initially,
\begin{align}
B_Y^s(k,t)=& \frac{\alpha \epsilon}{\sqrt{(1-\alpha)^2+4 \alpha \epsilon^2}}\left( \exp\left[ -  \dfrac{1+\alpha-\sqrt{(1-\alpha)^2+4 \alpha \epsilon^2}}{2 \alpha} \frac{k^2}{\sigma_Y}(t-t_i)\right] \right. \notag \\ 
&\left. -  \exp\left[ -  \dfrac{1+\alpha+\sqrt{(1-\alpha)^2+4 \alpha \epsilon^2}}{2 \alpha} \frac{k^2}{\sigma_Y}(t-t_i)\right] \right) B_D^s(k,t_i), \label{sol5} \\
B_D^s(k,t)=& \left( \frac{-1+\alpha+\sqrt{(1-\alpha)^2+4 \alpha \epsilon^2}}{2 \sqrt{(1-\alpha)^2+4 \alpha \epsilon^2}}  \exp\left[ -  \dfrac{1+\alpha-\sqrt{(1-\alpha)^2+4 \alpha \epsilon^2}}{2 \alpha} \frac{k^2}{\sigma_Y}(t-t_i)\right] \right. \notag \\ 
&\left. - \frac{-1+\alpha-\sqrt{(1-\alpha)^2+4 \alpha \epsilon^2}}{2 \sqrt{(1-\alpha)^2+4 \alpha \epsilon^2}}  \exp\left[ -  \dfrac{1+\alpha+\sqrt{(1-\alpha)^2+4 \alpha \epsilon^2}}{2 \alpha} \frac{k^2}{\sigma_Y}(t-t_i)\right] \right) B_D^s(k,t_i).  \label{sol6}
\end{align}
Thus, even if magnetogenesis only comes from the dark sector, nonvanishing visible magnetic fields 
are still produced from the magnetic field transfer between two sectors. 

The dark-to-visible transfer is a consequence of the difference of the 
effective electric conductivities in the basis ${\hat B}_Y$ 
and ${\hat B}_D$. 
As we discuss below, although both $\hat{B}$-fields dissipate through Eq.~(\ref{sol1}) and (\ref{sol2}), 
the incomplete cancellation between them lead to the linear growth of the visible field when 
$k^2 (t-t_i)/\sigma_Y \ll 1$ with a size that is inverse proportional to the conductivity. We have been 
deriving results by assuming particles charged either under the SM or dark U$(1)$ but not both.
The result can alternatively be understood qualitatively in a different charge basis. 
From the original basis in the Lagrangian of Eq.~\eqref{lag}, we can go to the 
basis without gauge kinetic mixing but with mixed currents that are charged under both dark and visible
U(1) symmetries. 
Then the dark magnetic fields are associated with the current charged under both the dark and visible U(1), 
and the nonzero visible U(1) charge carried by the current in this basis 
sources the visible magnetic fields.

\begin{figure}
\begin{center}
\includegraphics[height=8cm]{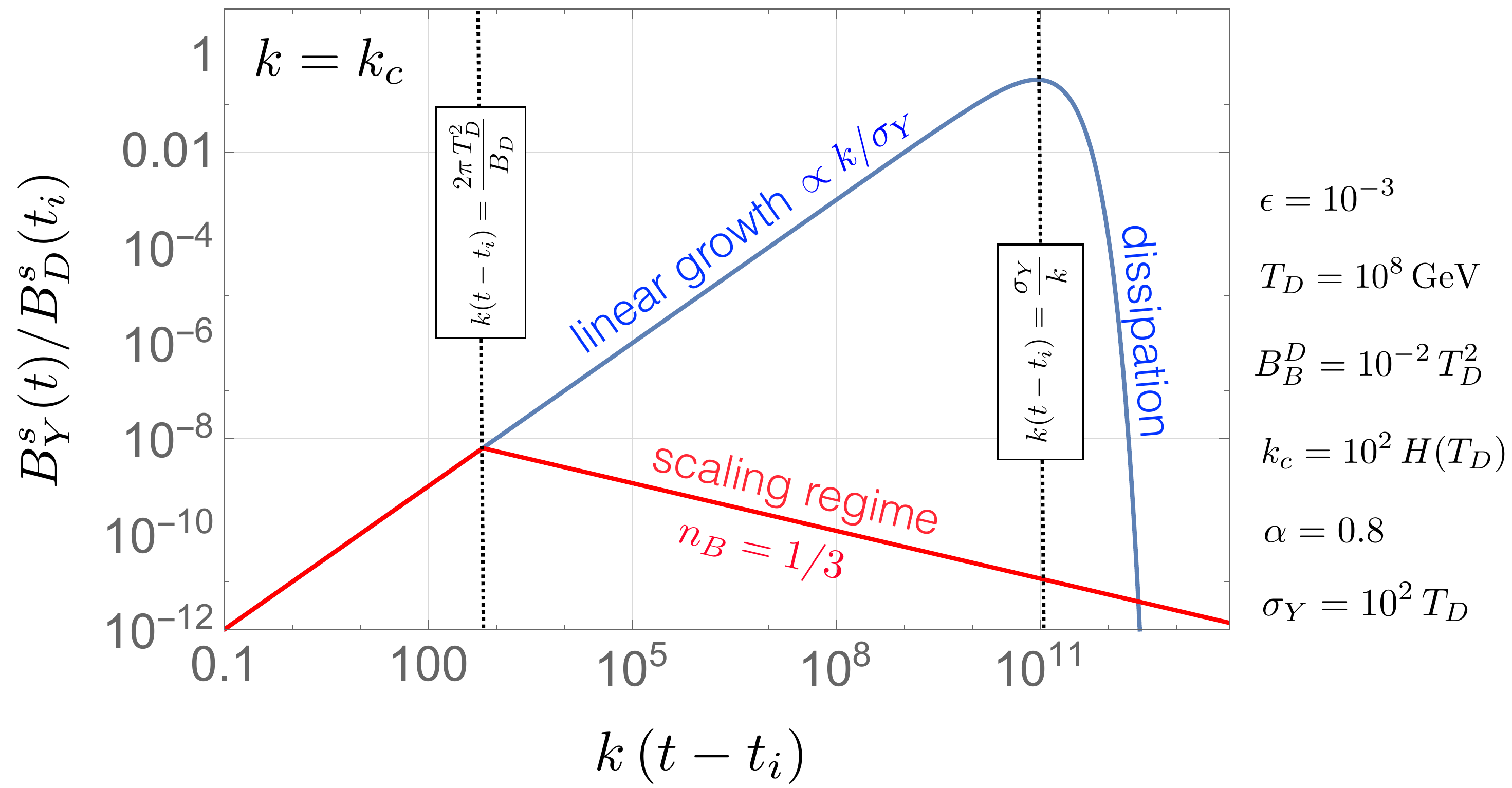}
\caption{An illustration of the evolution of mean visible $B$-field strength from the dark $B$-field transfer as a function of conformal time that is divided by the coherence length of the dark $B$-field ($k_c=2\pi/\lambda_c$). Blue curve shows the growth and dissipation of the visible $B$-field, $B^{s}_Y$, if the co-moving eddy scale of the turbulence, $(t-t_i)B_D(t_i)/T_D^2$, is much smaller than the coherence length $2\pi/k_c$ of the dark $B$-field. However, before $B^{s}_Y$ grows to its maximum size $\epsilon B^{s}_{D}$ permitted by the kinetic mixing, the fluid velocity cannot be ignored and turbulence becomes important. 
Then if there is no dynamo amplification, the
$B$-field decays following the scaling law discussed in Sec.~\ref{sec4} and is shown by the red curve in the plot.
}\label{fig:Btransfer}
\end{center}
\end{figure}

We show an example of the visible $B$-field evolution in Fig.~\ref{fig:Btransfer}. We are interested in scenarios with photon mixing $\epsilon\ll 1$ and conductivity ratio $\alpha\sim1$. At early times, $k^2 (t-t_i)/\sigma_Y \ll 1$, the visible field grow linearly,
\begin{equation}
B_Y^s(k,t)\simeq \frac{\epsilon k^2 }{\sigma_Y} (t-t_i) B_D^s(k,t_i) 
\label{startofdiss}
\end{equation}
while after $k^2 (t-t_i)/\sigma_Y >1$ the field decays exponentially $\propto \exp[-k^2 (t-t_i)/\sigma_Y]$
due to the usual diffusion effects (blue curve). 
The change from growth to decay will occur at time $t-t_i \approx \sigma_Y/k^2$,
once the visible magnetic field has grown to
$B_Y^s(k,t) \approx \epsilon B_D^s(k,t_i)$.
However, in Sec.~\ref{sec4} we will see that our assumption $v \approx 0$ breaks 
down before the dissipative regime can start for the case with relatively small coherence
length, and we have to use the full MHD solution 
that takes the fluid velocity into account (red curve).
The efficiency of the transfer, $\epsilon k^2 (t-t_i)/\sigma_Y$, 
is the same for both helicity modes. Since there is 
no transfer to the opposite helicity mode, 
the helicity-to-energy ratio $(|B^+(k,t)|^2-|B^-(k,t)|^2)/(|B^+(k,t)|^2+|B^-(k,t)|^2)$ for {\it each} $k$ mode 
is conserved during the dark to visible $B$-field transfer. 
However, unless the magnetic field spectrum is dominated by a single $k$ mode, the {\it total} helicity-to-energy 
ratio in the visible magnetic fields obtained by integrating over all $k$ modes may differ from
that in the dark magnetic fields if the helicity-to-energy ratio is $k$-dependent. 
In the case of maximally helical fields, only one of $B^\pm$ is nonvanishing. 
Maximally helical $B_Y$ emerges
from maximally helical $B_D$ with identical polarizations,
independent of the spectrum since all the $k$ modes are maximally helical.

\subsection{Second stage: $v \ne 0$ \label{3b}}

Through the Lorentz force that acts on the charged particles in the fluids, 
velocity fields are eventually generated from the magnetic fields,  
and the fluid becomes turbulent.
At that stage, the standard MHD studies (without dark magnetic fields)
have shown that the magnetic fields evolve according to 
a scaling law that depends on whether there is an inverse 
cascade~\cite{Banerjee:2004df,Kahniashvili:2012uj,Frisch1975,Pouquet1976}, 
direct cascade~\cite{Biskamp,Banerjee:2004df,Jedamzik2011,Kahniashvili:2012uj}, 
or inverse transfer~\cite{Brandenburg:2014mwa,Zrake:2014mta,Reppin:2017uud,Brandenburg:2017neh}.
In any case, as a first approximation, the magnetic fields are described by the comoving
field strength $B_c(t)$ at the coherence length $\lambda_c(t)$ or 
the peak scale $k(t) = 2\pi/\lambda_c(t)$ in the conformal frame, and they evolve as
\begin{equation}
B_c(t)  \propto t^{-n_B},  
\quad \lambda_c(t)\propto t^{n_\lambda}. 
\label{scalinglaws}
\end{equation}
Here $n_B$ and $n_\lambda$ are positive constants, 
which are determined by the helicity of the magnetic fields and properties of the turbulence~\cite{Reppin:2017uud,Brandenburg:2017neh}. 
Supposing that (i) the equilibration of the magnetic fields and velocity fields,
$v \sim B_c/\sqrt{\rho_c} $, where $\rho_c \sim T_c^4$ denotes the comoving fluid energy density 
($T_c$ is the temperature when the scale factor $a=1$), 
is established when the system enters the scaling regime\footnote{This is equivalent to the statement that
the velocity fields are amplified up to the Alfv\'en velocity $v_A = B_{\rm phys}/\sqrt{\rho}$ with $\rho$ being the physical (charged) fluid energy density.},
and (ii) the coherence length is determined by the eddy scale of the turbulence, 
$\lambda_c \sim v t \propto B_c t$, we have $n_B = 1- n_\lambda$,
which is also seen in the MHD simulations.
Analytical explanations of the scaling behavior, such as those given in
Refs.~\cite{Banerjee:2004df,Kahniashvili:2012uj,Jedamzik2011,Olesen:1996ts}, 
suggest that these exponents are insensitive to the values of MHD parameters. 

We assume Eq.~(\ref{sfapp}) is satisfied and take the single fluid approximation. This means the coherence lengths in two sectors are determined by the same eddy turnover scale and velocity field, $\lambda_{Y,\hat{Y}}\simeq\lambda_{D,\hat{D}}\simeq v\,t$. In our setup, when $B_D \gg B_Y$ in the un-hatted
frame, we have the following relations in the hatted frame,
\begin{equation}\left\{ \begin{array}{ll}
{\hat B}_D \simeq {\hat B}_Y \simeq B_D, &  \text{for} \quad \alpha \simeq 1, \\
{\hat B}_D \simeq   B_D,  \quad {\hat B}_Y \simeq \dfrac{\epsilon^2 \alpha}{(1-\alpha)^2} B_D, &    \text{for} \quad \alpha \ll 1, \\
{\hat B}_D \simeq \dfrac{\epsilon^2 \alpha}{(1-\alpha)^2} B_D,  \quad {\hat B}_Y \simeq   B_D, &    \text{for} \quad \alpha \gg 1. 
\end{array}
\right.  
\end{equation}
For $\alpha \simeq 1$, both of the hatted magnetic fields drive the plasma velocity,  $v \simeq B_D/T_c^2 = B_{D, {\rm phys}}/T^2$ 
with ${\hat B}_D \simeq {\hat B}_Y \simeq B_D$, and 
the hatted magnetic fields individually evolve according to the scaling laws, Eq.~\eqref{scalinglaws}.
For $\alpha \ll 1$, the velocity fields are driven by the ${\hat B}_{D}$ fields, and the
${\hat B}_{D}$ field strength as well as the coherence length evolve according to the scaling law. 
Similarly, for $\alpha \gg 1$, the velocity fields are driven by the ${\hat B}_{Y}$ fields, and 
${\hat B}_Y$ field strength as well as the coherence length evolve according to the scaling law. 
In these latter two cases, since the coherence length for the 
weaker hatted field,
${\hat B}_{Y}$ for $\alpha \ll 1$ and ${\hat B}_{D}$ for $\alpha \gg 1$, is determined by the same
eddy turnover scale, which is the same as the coherence length of 
the dominant hatted field, evolving with the scaling exponent $n_\lambda$,  
we expect that both magnetic fields ${\hat  B}_Y$ and ${\hat B}_D$ 
evolve or decay with a scaling law of the same exponent once the turbulence is fully developed 
 (see Appendix~\ref{appb} for more detailed explanation).
Since the visible and dark magnetic fields in the original basis (Eq.~\eqref{lag}) 
are linear combination of those with the basis  
${\hat B}_Y$ and ${\hat B}_D$ (Eqs.~\eqref{bas1} and \eqref{bas2}), 
the ratio between the magnitude of dark and visible magnetic fields are fixed 
during turbulent evolution, and there is no further field transfer at the order of our 
approximation.

In scenarios of magnetogenesis such as those from the first order phase 
transitions, the system enters the scaling regime before the 
dissipation starts to erase the magnetic fields exponentially, 
$k_c^2 (t-t_i)/\sigma_Y <1$, 
with $k_c$ being the characteristic scale of the magnetic fields. 
In the next section we discuss a concrete setup
and evaluate the visible magnetic field surviving until today.

Before we end the section, we want to comment on a possible amplification of $B_Y$. 
It should be noticed that there is no numerical study yet for the case when $\rho_{\hat D}\gsim\rho_{v}\gg\rho_{\hat Y}$, where $\rho_{\hat{D},\hat{Y}}$ and $\rho_v$ are the energy densities in the hatted $B$-fields and fluid velocity field.
It remains as a possibility that the weaker ${\hat B}_Y$ field (for $\alpha \ll 1$)
can experience dynamo-amplification and/or will enter the scaling regime at some time after
the stronger ${\hat  B}_D$  field starts scaling
(and vice versa for $\alpha \gg 1$). 
The two $\hat{B}$ fields may therefore evolve with different scaling exponents for some time, which may result in
additional amplification of the visible $B_Y$ fields. A quantitative estimation of such an amplification requires detailed numerical
simulations. Here we take a conservative position where we assume that such amplifications of visible $B$-field are negligible.

\section{Estimate of present intergalactic magnetic field \label{sec4}}

Now we evaluate the strength and coherence length of the present visible magnetic fields 
in a concrete setup.
Here we assume that the dark magnetic fields are generated at an early time
before the electroweak phase transition and transferred to the SM hyper U(1) magnetic fields. 
Then the hyper magnetic fields smoothly turn into the (electro)magnetic fields at the electroweak 
symmetry breaking without any decay or amplification, which evolve according to the scaling law and remain until today. 
Let us write the temperature at dark magnetogenesis as
$T(t_i)=T_D$ and parameterize
the typical momenta (or the inverse of the coherence length) 
of the dark magnetic fields as $k_c \sim \gamma H_D$.  
Here $H_D = \sqrt{8\pi^3 g_*/90}\,T_D^2/m_{\rm Pl}\simeq 1.66\,g_*^{1/2} T_D^2/m_{\rm Pl}$ is the Hubble parameter at the dark magnetogenesis with $g_*$ being the number of relativistic degrees of freedom.\footnote{We assume $g_*(T_D)=213.5$ when presenting result. The number is the sum of SM degrees of freedom above the electroweak scale ($106.75$) and the number from dark sector (106.75) that contains a mirror copy of SM particles with the same SM temperature.}   $\gamma$ 
is the ratio between the Hubble radius and initial magnetic field coherence length, which we take as a 
free parameter that parameterizes the magnetogenesis models. 
If we specify 
a magnetogenesis model, $\gamma$ 
can be obtained {\it e.g.}, from numerical simulations. For instance, $\gamma\sim10^2$~\cite{Durrer:2013pga} 
if the dark magnetogenesis comes from a first order phase transition,
and the initial coherence length of the magnetic fields is the order of the size of the largest bubbles at coalescence.

The magnetic field strength at this time is denoted by $B_D(t_i)$ 
and has energy density $\sim (B_D(t_i))^2$, which
can be comparable to the energy density of the thermal fluids $\rho=\zeta T_D^4$ with $\zeta = \pi^2 g_*/30$.
At first the dark magnetic fields evolve adiabatically 
except for the slight decay due to dissipation. 
The first stage terminates when the coherence length $2\pi/k_c$ 
is caught up by the eddy turnover scale $v \Delta t_{\rm s}$ with $v \approx v_A\approx \zeta^{-1/2} B_D(t_i)/T_D^2$, supposing that the velocity fields gets equilibrated to the magnetic fields at 
a sufficiently earlier time. 
$\Delta t_{\rm s}$ is the time interval of the first stage, and we take the scale factor to be
$a=1$ at $t=t_i$ to write it as
\begin{equation}
\Delta t_{\rm s} \sim \frac{2 \pi \zeta^{1/2} T_D^2}{k_c B_D(t_i)} \sim \frac{\sqrt{6\pi}}{2\gamma} \frac{m_{\rm Pl}}{B_D(t_i)}.
\label{Deltats}
\end{equation}
The dissipative evolution starts at $t-t_i \simeq \sigma_Y/k_c^2 \simeq 3\sigma_Y m_{\rm Pl}^2/8 \pi\gamma^2 \zeta T_D^4$
. For a sufficiently small $\gamma$, \begin{equation}
\gamma \lesssim  \frac{\sqrt{3}\sigma_Y  }{2 (2 \pi)^{3/2} \zeta T_D}\left(\frac{B_D(t_i)}{T_D^2}\right)\left(\frac{m_{\rm Pl}}{T_D}\right) 
, \label{gamma}
\end{equation}
the first stage ends before the system reaches the regime of dissipative evolution ($\Delta t_{\rm s}\ll t-t_i$). Since we are interested in an efficient magnetogenesis $\sim B_D(t_D)/T_D^2 \sim 10^{-2}$   and focus on scenarios with $T_D\leq 10^{14}$ GeV
for the single fluid approximation, the inequality becomes $\gamma \lsim  10^2$ after taking $\sigma_Y\simeq100\,T_D$ from Eq.~(\ref{eq:conduc}). The equality can be naturally satisfied, {\it e.g.}, in magnetogenesis from the first order phase transition 
at $T_D \ll 10^{14}$ GeV
that has typical $\gamma  \simeq 10^2$~\cite{Durrer:2013pga}. We assume the inequality holds for the following derivation.

Now let us evaluate the properties of the visible magnetic fields at the present epoch.
In order to compare them with the observations here we move to the physical frame.
Combining Eqs.~(\ref{startofdiss}), (\ref{Deltats}) and the relations
$k_c = \gamma H_D$, the coherence length at the time of dark magnetogenesis $\lambda_Y = 2 \pi/k_c$, 
 and the Hubble redshift factor $(H_D \Delta t_s)^{-2}$
of the magnetic field strength (note that $t$ denotes conformal time),
the physical visible magnetic field strength and coherence length when the first stage ends and the system enters the 
scaling regime {($t=t_{\rm s}$)} are
\begin{align}
B_{Y,{\rm phys}} (t_s) & \simeq \frac{\epsilon k_c^2}{\sigma_Y} \Delta t_{\rm s}B_D(t_i)  (H_D \Delta t_s)^{-2}   \simeq  \frac{\epsilon \gamma^3}{2 \pi \zeta^{1/2}} \frac{H_D T_D^2}{\sigma_Y} \left(\frac{B_D(t_i)}{T_D^2}\right)^2 \notag \\
&= 3.8 \times 10^{3} \ {\rm GeV}^2  \left(\frac{\epsilon}{10^{-1}}\right) \left(\frac{\sigma_Y}{10^2 T_D}\right)^{-1} \left(\frac{\gamma}{10^2}\right)^3 \left( \frac{B_D(t_i)/T_D^2}{0.01}\right)^2 \left(\frac{T_D}{10^{8} {\rm GeV}}\right)^3 \notag \\
& = 5.6 \times 10^{22} \ {\rm G} \left(\frac{\epsilon}{10^{-1}}\right) \left(\frac{\sigma_Y}{10^2 T_D}\right)^{-1} \left(\frac{\gamma}{10^2}\right)^3 \left( \frac{B_D(t_i)/T_D^2}{0.01}\right)^2 \left(\frac{T_D}{10^{8} {\rm GeV}}\right)^3, \label{mags} \\
\lambda_{Y, {\rm phys}} (t_{\rm s}) & \simeq \frac{2\pi}{k_c} (H_D \Delta t_s)  \simeq \sqrt{\frac{3}{8\pi}}\left(\frac{2 \pi}{\gamma}\right)^2 \frac{m_{\rm Pl}}{B_D(t_i)} \notag \\
& \simeq 1.6 \times 10^2 \ {\rm GeV}  \left(\frac{\gamma}{10^2}\right)^{-2} \left(\frac{B_D(t_i)/T_D^2}{0.01} \right)^{-1} \left(\frac{T_D}{10^8 {\rm GeV}}\right)^{-2} \notag \\
&\simeq 1.1 \times 10^{-36} {\rm Mpc}  \left(\frac{\gamma}{10^2}\right)^{-2} \left(\frac{B_D(t_i)/T_D^2}{0.01} \right)^{-1} \left(\frac{T_D}{10^8 {\rm GeV}}\right)^{-2}.  \label{coh}
\end{align}

Here we have taken into account the redshift from the magnetic field generation to the onset of 
the scaling law using,
\begin{equation}
H_D \Delta t_{\rm s} \simeq \frac{2\pi \zeta^{1/2}T_D^2}{\gamma B_D(t_i)}, 
\end{equation}
which also gives the temperature at the onset of the scaling evolution 
(when fluid velocity cannot be ignored),
\begin{equation}
T_{\rm s} \simeq \frac{T_D}{H_D \Delta t_{\rm s}} = \frac{\gamma B_D(t_i)}{2 \pi \zeta^{1/2} T_D} = 1.9 \times 10^6 {\rm GeV} \left(\frac{\gamma}{10^2}\right) \left(\frac{B_D(t_D)/T_D^2}{0.01}\right)\left(\frac{T_D}{10^8{\rm GeV}}\right). \label{temps}
\end{equation}
Note that this expression applies only
for $B_D(t_i)<2 \pi T_D^2/\gamma$. For $B_D(t_i)\gsim2 \pi  \zeta^{1/2} T_D^2/\gamma$, magnetic fields will be entering the scaling regime
in a Hubble time but we do not consider such cases. 

Assuming that the visible magnetic fields evolve according to the scaling law 
without experiencing significant dynamo amplifications
until recombination 
and afterwards evolve adiabatically until today, we can estimate the present strength and coherence length of the intergalactic magnetic fields using the scalings,
\begin{equation}
B_{{\rm phys}}^0=\left(\frac{a(t_{\rm s})}{a_0}\right)^{2} \left(\frac{t_{\rm s}}{t_{\rm rec}}\right)^{n_B} B_{Y,{\rm phys}}(t_{\rm s}), \quad \lambda_{\rm phys}^0 = \left(\frac{a(t_{\rm s})}{a_0}\right)^{-1} \left(\frac{t_{\rm s}}{t_{\rm rec}}\right)^{-n_\lambda} \lambda_{Y, {\rm phys}}(t_{\rm s}), \label{scsc}
\end{equation}
where $B_{\rm phys}^0$ and $\lambda_{\rm phys}^0 $ are the physical magnetic field strength and coherence strength today,
respectively, and $t_{\rm rec}$ is the conformal time at recombination. 
Since we are interested in getting maximal $B$-field and coherence length, we assume the dark $B$-field is generated with maximal helicity, in which case 
the exponent of the scaling exponent is known as
$n_B=1/3$ and $n_\lambda = 2/3$~\cite{Banerjee:2004df,Kahniashvili:2012uj}. 
(See also Appendix~\ref{appb}.) For example, maximally helical $B$-field can be generated by 
pseudo scalar inflation~\cite{Garretson:1992vt,Anber:2006xt} and 
chiral instability~\cite{Joyce:1997uy,Tashiro:2012mf} (see also Refs.~\cite{Jimenez:2017cdr,Kamada:2018tcs}).  
Besides taking it as an assumption for generating large magnetic fields, the existence of helical intergalactic magnetic fields may even be indicated by data from the parity-violating correlations of the diffuse gamma ray flux~\cite{Tashiro:2013bxa,Tashiro:2013ita,Tashiro:2014gfa,Chen:2014qva}. This gives a motivation to study the maximally-helical scenario, while our derivations can be easily adapted to different $(n_B,n_{\lambda})$ assumptions.

From Eqs.~\eqref{mags}, \eqref{temps}, and \eqref{scsc} 
we obtain the present magnetic field strength and coherence length in terms of the temperature and field strength 
at the dark magnetogenesis as, 
\begin{align}
B_{\rm phys}^0 &\simeq 2.8 \times 10^{-22} {\rm G} \left( \frac{\gamma}{10^2}\right)^{2/3} \left(\frac{\epsilon}{0.1}\right) \left(\frac{\sigma_Y}{100 T_d}\right)^{-1}    \left(\frac{B_D(t_i)/T_D^2}{0.01}\right)^{-1/3} \left(\frac{T_D}{10^{8}{\rm GeV}}\right)^{2/3}, \label{fbs}\\
\lambda_{\rm phys}^0 & \simeq 7.5 \times 10^{-6} {\rm Mpc} \left(\frac{\gamma}{10^2}\right)^{-1/3} \left( \frac{B_D(t_i)/T_D^2}{0.01}\right)^{2/3}  \left(\frac{T_D}{10^{8}{\rm GeV}}\right)^{-1/3}. \label{fbc}
\end{align}

\begin{figure}
\begin{center}
\includegraphics[height=10cm]{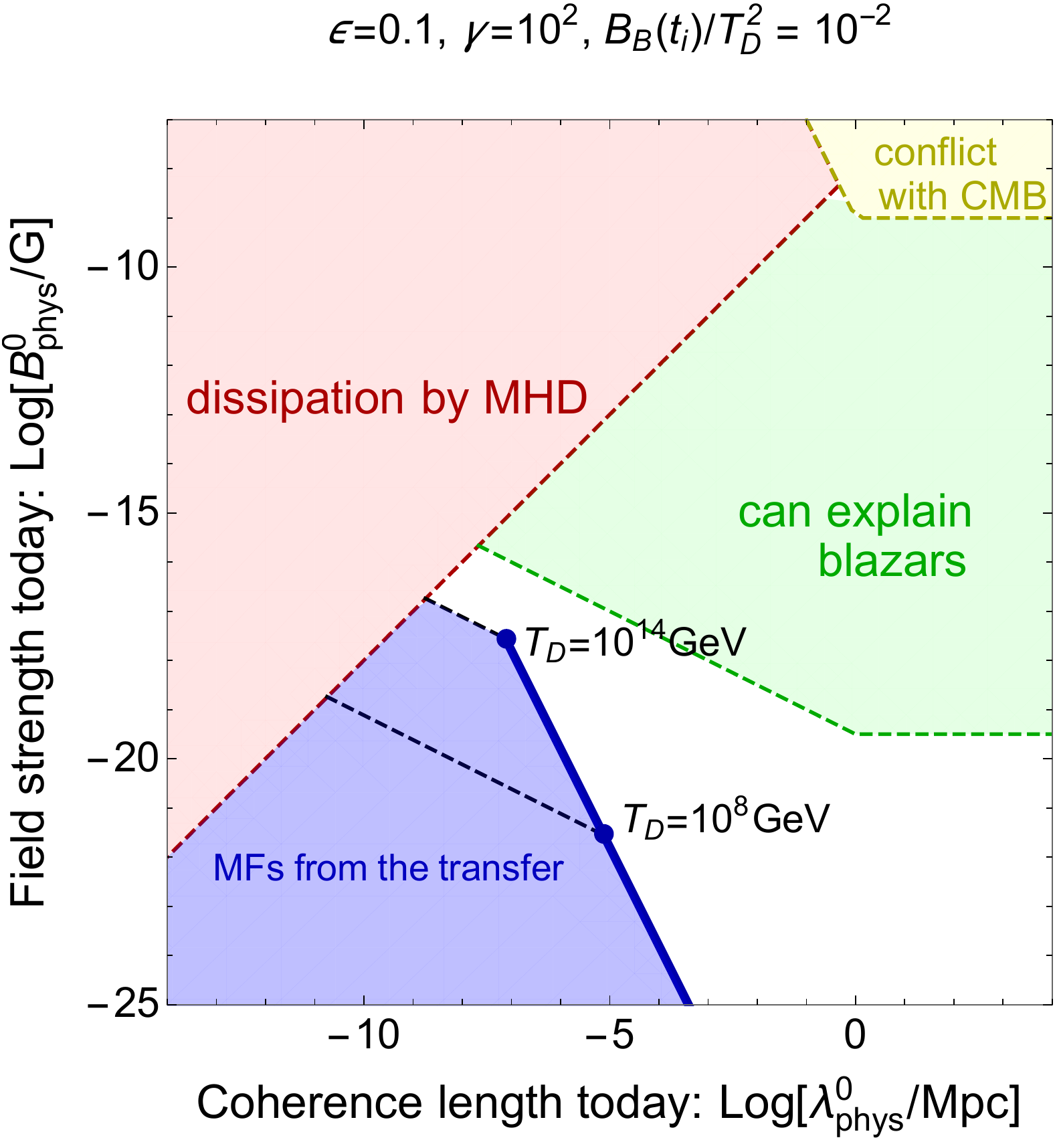}
\caption{
The magnetic field properties today for the maximally helical case.
The blue thick solid line represents the magnetic field properties assuming  
$T_D<10^{14}$ GeV and constant scaling exponents.
Blue shaded region represents the parameter space if
we take into account the possible change of the scaling exponents after dark U(1) symmetry breaking. 
The black dashed lines
represent how the final magnetic field properties differ  for $T_D=10^8$~GeV (lower line)
and $T_D=10^{14}$~GeV (upper line) if the scaling exponents change
at the dark U(1) symmetry breaking while keeping the helicity conservation. 
The other parameters are fixed as $\epsilon=0.1, \gamma= 10^2$ and $B_D(t_i)/T_D^2 = 10^{-2}$. 
Green region is the region where the blazar observations are explained~\cite{Finke:2015ona}. 
The region inconsistent with the MHD evolution and the one in conflict with CMB observations 
are depicted in the red colored region and yellow colored region, respectively. 
}\label{fig:today}
\end{center}
\end{figure}

In Fig.~\ref{fig:today}, the field strength and coherence length of the present magnetic fields 
for different choices of $T_D$ 
are depicted with the blue thick solid line.
We can see that such fields are too weak and incoherent
to explain the observed deficit of secondary GeV cascade
photons from blazars even if we take relatively extreme parameters such as 
$T_D \simeq 10^{14}$~GeV (comparing to Eq.~(\ref{sfapp})) and a large $\gamma$. 
The main reason for the weak field strength is that the transfer efficiency factor  
$\epsilon k_c^2 \Delta t_{\rm s}/\sigma_Y$ is very small, 
$\sim 1.0\times 10^{-7}(\epsilon/0.1) (\gamma/10^2) ((B_D(t_i)/T_D^2)/0.01)^{-1} (T_D/10^8{\rm GeV})$
due to the large electrical conductivity ($k_c/\sigma_Y\sim H_D/T_D\ll 1$). One might think that taking 
$\gamma T_D>10^{16}$~GeV and $B_D(t_i)/T_D^2 < 10^{-2}$ can increase the size of 
$B^0_{\text{phys}}$ in \eqref{fbs}. But this violates the condition in Eq.~\eqref{gamma}, and the exponential 
dissipative decay happens too early and eliminate the magnetic fields 
before the turbulent plasma develops, so Eqs.~\eqref{fbs} and \eqref{fbc} do not apply. Moreover, since magnetic fields 
decay faster in the nonhelical or partially helical case, the maximally helical case we study should give the largest 
visible $B$-field.
 
In order to avoid all possible collider and cosmological constraints, we have in mind that the dark U$(1)_D$ 
symmetry breaks down at a high temperature {\it e.g.}, above the electroweak symmetry breaking.
It is not quite clear if the scaling relation with the exponents in Eq.~(\ref{scsc})
holds after the U$(1)_D$ symmetry breaks. 
When deriving Eq.~\eqref{fbs} and \eqref{fbc} with the exponents $n_B=1/3$ and $n_\lambda = 2/3$, 
we implicitly assume 
helicity conservation for each of the ${\hat B}$ fields, $\lambda {\hat B}_{Y,B}^2/2\pi$=const., 
with the same coherence length that is comparable to the eddy turnover scale $\lambda\sim vt$, 
where the velocity fields are in equilibrium with the {\it dark} magnetic fields  
$v\sim v_A \sim B_{D,{\rm phys}}/\sqrt{\rho}$~\cite{Fujita:2016igl,Kamada:2016eeb}. (See also Appendix~\ref{appb}.)
However, these assumptions do not hold after dark U$(1)_D$ symmetry breaking. 
The Alfv\'en velocity evaluated with the $B_Y$ field after U$(1)_D$ symmetry breaking 
will be much weaker than the velocity fields we used to derive Eqs.~\eqref{fbs} and \eqref{fbc}.  
Thus we also expect that the 
eddy turnover scale to be shorter after U$(1)_D$ breaking. 
Consequently the coherence length of the $B_Y$ fields will be
smaller than the estimate in Eq.~(\ref{fbc}), which is the largest possible coherence length 
that can be achieved in this scenario. 

The actual coherence length as well as the $B_Y$ field strength depend on the decay of the velocity fields
and can be estimated as follows.
Suppose that the velocity fields cannot become smaller than the Alfv\'en velocity for the $B_Y$ field, the $B_Y$ 
field strength and the minimal coherence length satisfy the relation 
$\lambda_{\rm phys}^0/{\rm pc} \simeq B_{\rm phys}^0/(10^{-14} {\rm G})$~\cite{Banerjee:2004df,Durrer:2013pga} (see also Appendix \ref{appc}),
which is the red dashed line in Fig.~\ref{fig:today}. 
Therefore, if there is no additional dynamo-amplification of the $B_Y$ fields, the actual magnetic field properties will 
lie between the red dashed line ($\lambda_{\rm phys}^0/{\rm pc} \simeq B_{\rm phys}^0/(10^{-14} {\rm G})$) 
and the blue solid line (Eq.~\eqref{fbs} and \eqref{fbc}) in Fig.~\ref{fig:today}.

How can we determine the magnetic field properties further? 
We can take advantage of helicity conservation. 
Since we consider the maximally helical case, the helicity density
$h=\lambda_{\rm phys}^0 (B_{\rm phys}^{0})^2/2\pi$ is conserved.
Thus we expect that the final magnetic field properties lie on the black dashed line 
shown in Fig.~\ref{fig:today} 
with respect to the temperature at the dark magnetogenesis $T_D$. 
In summary, we conclude that the magnetic field properties today in this scenario lie in the blue shaded region 
in Fig.~\ref{fig:today}. 
We can see that even taking into account the possible faster
decay of the velocity fields after the dark U(1) symmetry breaking, 
the present 
field strength and coherence scale
of the magnetic fields are below the lower bound of the numbers that can explain the 
blazar observations. 
Therefore, in our conservative estimate, it is 
not possible to explain the blazar observation by the magnetic fields
transferred from the dark U(1) fields. 
However we do not exclude the possibility where 
the dynamo-amplification works when the scaling regime starts or the dark U$(1)_D$ symmetry breaks down, 
so that the amplified magnetic fields may still explain the observation.

Before closing our discussion, let us mention that we have not considered another issue on
the dark U(1) symmetry breaking. When the symmetry breaks down, the dark U(1) gauge boson gets massive and
the dark magnetic fields are confined to cosmic strings. The implications for the visible magnetic
fields will then depend on when the dark U(1) symmetry breaking occurs.

\section{Summary \label{sec5}}

In this article, we have examined how dark magnetic fields can be transferred to the visible 
magnetic fields through the gauge kinetic mixing. 
We have considered the system with the Lagrangian Eq.~\eqref{lag} 
where there are independent dark and visible U(1) currents in the basis 
with gauge kinetic mixing. 
We have found that in such  a system 
the visible magnetic fields emerge
due to the transfer from the dark magnetic fields in thermal fluids through the gauge kinetic mixing, 
when the velocity fields are small. 
The efficiency is suppressed by the gauge kinetic mixing parameter $\epsilon$ as well as 
the large electric conductivity $\sigma_Y$ and the duration of the transfer. 
At some later time the velocity fields develop turbulence and the transfer terminates
when the magnetic fields enter the scaling regime. 
The ratio between the visible and dark magnetic field strength is fixed at that time. 
Due to the shortness of the duration for the system to enter the scaling regime, 
the visible-to-dark magnetic field strength ratio is generally very small, say $\sim 10^{-7}$. 
As a result, it is not possible 
to explain the TeV blazar observations by the visible intergalactic magnetic fields
generated by this mechanism without further dynamo amplification. 

We have not considered the dark U(1) symmetry breaking in detail, which should be associated with 
a possible dynamo amplification of the visible magnetic fields as well as
the dark cosmic string production. 
We assume the symmetry breaking does not affect the abundance of relic visible magnetic fields. 
However, we do not exclude the possibility that this would change the strength of the 
visible magnetic fields. 
This is left for future work.

\section*{Acknowledgements}
The work of KK was supported by IBS under the project code, IBS-R018-D1,
and of KK and TV at ASU by the DOE under Grant No. DE-SC0013605. YT was 
supported in part by the National Science Foundation under grant PHY-1620074, 
and by the Maryland Center for Fundamental Physics.
KK also would like to thank the Maryland 
Center for Fundamental Physics for their kind hospitality where this work was initiated.

\appendix

\section{The case with particles charged under both U(1) symmetry \label{app1}}

In the main part of the present article, we focused on the dark-visible U(1) gauge system with 
visible and dark currents that are independent of each other as described by 
Eq.~\eqref{ohm1}. 
We now examine the case where there are particles charged 
under both dark and visible U(1) symmetry.\footnote{The case 
when particles charged under dark U(1) symmetry are absent is studied at Ref.~\cite{Choi:2018dqr}}.
This is the case {\it e.g.}, when U(1)$_{B-L}$ magnetic fields are generated. 
In this case, both dark and visible U(1) currents consist of the current of these particles
and are not independent.
Then the off-diagonal components of the electric conductivity 
need to be taken into account.
The discussion in this appendix also shows the validity of the treatment in the main part
that we neglect the off-diagonal components of the electric conductivity.

Compared to the discussion in the main article, 
Ohm's law in Eq.~\eqref{ohm1} is modified to 
\begin{align}
{\bm J}_Y&= \sigma_Y ({\bm E}_Y+{\bm v}\times {\bm B}_Y)+\sigma_{YB} ({\bm E}_D+{\bm v}\times {\bm B}_D),   \label{ohm2}\\
{\bm J}_D&= \sigma_D ({\bm E}_D+{\bm v}\times {\bm B}_D)+\sigma_{YB} ({\bm E}_Y+{\bm v}\times {\bm B}_Y),  \label{ohm3}
\end{align}
where the off-diagonal component of the electric conductivity $\sigma_{YB}$ is introduced, 
which we expect to satisfy $\sigma_{YB} \sim \sigma_Y \sim \sigma_D$. 
Eliminating the electric fields by using Eqs.~\eqref{ohm2} and \eqref{ohm3} and substituting them 
in the modified Maxwell's equations~\eqref{max1} and \eqref{max2}, 
we obtain the evolution equations for the magnetic fields as 
\begin{align}
{\dot {\bm B}}_Y &= \frac{1}{\sigma_D \sigma_Y - \sigma_{YB}^2} \left((\sigma_D-\epsilon \sigma_{YB}) {\bm \nabla}^2 {\bm B}_Y+(\epsilon \sigma_D-\sigma_{YB}){\bm \nabla}^2 {\bm B}_D\right)+{\bm \nabla} \times({\bm v} \times {\bm B}_Y) \notag \\
&=\frac{1}{{\hat \sigma}_Y} {\bm \nabla}^2 {\bm B}_Y+\frac{{\hat \epsilon}_Y}{{\hat \sigma}_Y} {\bm \nabla}^2 {\bm B}_D+{\bm \nabla} \times({\bm v} \times {\bm B}_Y),  \\
{\dot {\bm B}}_D 
&=\frac{1}{\sigma_D \sigma_Y - \sigma_{YB}^2} \left((\sigma_Y-\epsilon \sigma_{YB}) {\bm \nabla}^2 {\bm B}_D + (\epsilon \sigma_Y - \sigma_{YB}) {\bm \nabla}^2 {\bm B}_Y \right)+{\bm \nabla} \times({\bm v} \times {\bm B}_D) \notag \\
&=\frac{1}{{\hat \sigma}_D} {\bm \nabla}^2 {\bm B}_D+\frac{{\hat \epsilon}_D}{{\hat \sigma}_D} {\bm \nabla}^2 {\bm B}_Y+{\bm \nabla} \times({\bm v} \times {\bm B}_D), 
\end{align}
where
\begin{equation}
{\hat \sigma}_Y \equiv \frac{\sigma_D \sigma_Y - \sigma_{YB}^2}{\sigma_D-\epsilon \sigma_{YB}}, \quad {\hat \sigma}_D \equiv \frac{\sigma_D \sigma_Y - \sigma_{YB}^2}{\sigma_Y-\epsilon \sigma_{YB}}, \quad {\hat \epsilon}_Y \equiv \frac{\epsilon \sigma_D-\sigma_{YB}}{\sigma_D-\epsilon \sigma_{YB}}, \quad {\hat \epsilon}_D \equiv \frac{\epsilon \sigma_Y - \sigma_{YB}}{\sigma_Y-\epsilon \sigma_{YB}}.
\end{equation}
Then the evolution equations are decoupled as 
\begin{align}
{\dot {\hat {\bm B}}_Y} &= \dfrac{1+{\hat \alpha}-\sqrt{(1-{\hat \alpha})^2+4 {\hat \alpha} {\hat \epsilon}_Y{\hat \epsilon}_D}}{2 {\hat \alpha}}  \frac{{\bm \nabla}^2}{{\hat \sigma}_Y} {\hat {\bm B}}_Y +{\bm \nabla}\times ({\bm v} \times {\hat {\bm B}}_Y ), \\
{\dot {\hat {\bm B}}_D} &=  \dfrac{1+{\hat \alpha}+\sqrt{(1-{\hat \alpha})^2+4 {\hat \alpha} {\hat \epsilon}_Y{\hat \epsilon}_D}}{2 {\hat \alpha}}    \frac{{\bm \nabla}^2}{{\hat \sigma}_Y} {\hat {\bm B}}_D +{\bm \nabla}\times ({\bm v} \times {\hat {\bm B}}_D ), 
\end{align}
where 
\begin{equation}
{\hat \alpha} \equiv \frac{{\hat \sigma}_D}{{\hat \sigma}_Y} = \frac{\sigma_D-\epsilon \sigma_{YB}}{\sigma_Y-\epsilon \sigma_{YB}}, 
\end{equation}
and
\begin{align}
{\hat {\bm B}}_Y &\equiv -\dfrac{{\hat \epsilon}_D}{\sqrt{(1-{\hat \alpha})^2+4 {\hat \alpha} {\hat \epsilon}_Y{\hat \epsilon}_D}} {\bm B}_Y +\left( \dfrac{1}{2}-\dfrac{1-{\hat \alpha}}{2\sqrt{(1-{\hat \alpha})^2+4 {\hat \alpha} {\hat \epsilon}_Y {\hat \epsilon}_D}}\right) {\bm B}_D, \label{bas3}\\
 {\hat {\bm B}}_D &\equiv \dfrac{{\hat \epsilon}_D}{\sqrt{(1-{\hat \alpha})^2+4 {\hat \alpha} {\hat \epsilon}_Y {\hat \epsilon}_D}}{\bm B}_Y+\left( \dfrac{1}{2}+\dfrac{1-{\hat \alpha}}{2\sqrt{(1-{\hat \alpha})^2+4 {\hat \alpha}{\hat \epsilon}_Y {\hat \epsilon}_D}}\right) {\bm B}_D.  \label{bas4}
\end{align}
From this point on, we can take over the discussion from the main part of this article. Then we
conclude that:
\begin{enumerate}
\item When the velocity fields are negligible for the evolution of magnetic fields, {\it e.g.}, 
just after magnetogenesis, the magnetic field transfer occurs and the visible magnetic fields evolve as
\begin{equation}
B_Y^s (k,t) = \frac{{\hat \epsilon}_Y k^2}{{\hat \sigma}_Y}(t-t_i) B^s_D(k, t_i). 
\end{equation}
Since we expect $\sigma_{YB} \simeq \sigma_D \simeq \sigma_Y (\simeq {\hat \sigma}_Y)$, we have 
${\hat \epsilon}_Y \simeq -1$ 
and the sign of the visible magnetic field strength is opposite to the dark magnetic field strength. 
Note that if the off-diagonal component $\sigma_{YB}$ is suppressed by a factor of $\epsilon$, 
the result is the same to the one obtained in the main part of this article 
unless a fine-tuning 
${\hat \epsilon}_Y = 0$ or ${\hat \epsilon}_D = 0$, namely, $\epsilon \sigma_D = \sigma_{YB}$ or 
$\epsilon \sigma_Y = \sigma_{YB}$ is realized. 
Thus we can safely neglect the off-diagonal components of the electric conductivity
in the discussion of the main part.
\item When the system enters the scaling regime after the velocity fields fully develop, 
there will not be any further magnetic field transfer 
as long as both of the magnetic field spectra have the peaks at the same momentum 
and the scaling exponents are exactly the same.
\end{enumerate}

\section{The scaling laws for two fields \label{appb}}
Here we explain why the exponents for the scaling laws for the hatted fields are the same, 
argued in Sec.~\ref{3b}. 

\subsection{Maximally helical case}
In the maximally helical case, the reason why the exponents of two hatted 
fields ($\hat{B}_{B}$ and $\hat{B}_Y$ in our discussion) become the same can be explained as follows. 
In the conformal frame, we have the helicity conservations 
for both fields (denoted as $\hat{B}_{D}$ for the dominant field and $\hat{B}_S$
for the weaker field),
\begin{equation}
\lambda_D (t) {\hat B}_D^2 (t) = {\rm const}, \quad \lambda_S (t) {\hat B}_S^2 (t) = {\rm const}. 
\end{equation}
In the single fluid approximation, the coherence length are determined by the 
eddy turnover scale common to both fields, 
\begin{equation}
\lambda_D (t) \simeq \lambda_S (t) \simeq v(t)  t. 
\end{equation}
Supposing that the velocity fields are equilibrated to the dominant field, 
\begin{equation}
v(t)  \simeq \frac{{\hat B}_D(t) }{\sqrt{\rho}}, 
\end{equation}
where $\rho$ is the comoving energy density that is a constant. 
Then we first obtain the relations for the dominant field,
\begin{equation}
\lambda_D(t)  {\hat B}_D^2(t)  = {\rm const}, \quad \lambda_D (t) \simeq \frac{{\hat B}_D(t) }{\sqrt{\rho}} t, 
\end{equation}
which yeild
\begin{equation}
{\hat B}_D(t)  \propto t^{-1/3}, \quad \lambda_D(t)  \propto t^{2/3}. 
\end{equation}
Then we have 
$\lambda_S (t) \propto t^{2/3}$  for the weaker field. From the helicity conservation, 
$ \lambda_S(t)  {\hat B}_S^2 (t) = {\rm const}.$, 
we obtain
\begin{equation}
{\hat B}_S (t) \propto t^{-1/3}. 
\end{equation}
Thus the exponents of the scaling laws for both fields are the same
if both of them are maximally helical. 

\subsection{Nonhelical case with direct cascade}

The direct cascade for the nonhelical magnetic fields can be derived as follows. 
Suppose that the initial MF spectra are written as with the common exponents $n_s$ 
\begin{equation}
{\hat B}_D(k) = {\hat B}_D^0 \left(\frac{k}{k_0}\right)^{n_s},\quad {\hat B}_S(k) = {\hat B}_S^0 \left(\frac{k}{k_0}\right)^{n_s}, \quad \text{for} \quad k<k_0,
\end{equation}
with vanishing power at $k>k_0$, 
where ${\hat B}_D^0$ and ${\hat B}_S^0$ are understood as the typical field strength. 
If the velocity fields just erase smaller scale powers than the eddy turnover scale 
without amplifying the power at larger scales, 
the spectra are expressed as\begin{equation}
{\hat B}_D(k,t) = {\hat B}_D^0 \left(\frac{k}{k_0}\right)^{n_s},\quad {\hat B}_S(k,t) = {\hat B}_S^0 \left(\frac{k}{k_0}\right)^{n_s}, \quad \text{for} \quad k<\frac{2\pi}{v(t)  t}.
\end{equation}
Then the typical field strengths are given by 
\begin{equation}
{\hat B}_D(t) = {\hat B}_D^0\left(\frac{ 2 \pi }{k_0 v(t)  t}\right)^{n_s}, \quad {\hat B}_S(t) = {\hat B}_S^0\left(\frac{ 2 \pi }{k_0 v(t)  t}\right)^{n_s}, \label{d9}
\end{equation}
and the coherence lengths are given by the eddy turnover scale, 
\begin{equation}
\lambda_D (t) \simeq \lambda_S (t) \simeq v(t)t.  \label{d10}
\end{equation}
Again, supposing that the velocity fields are equilibrated to the dominant field, 
we can write 
\begin{equation}
v(t)  \simeq \frac{{\hat B}_D(t)}{\sqrt{\rho}}. \label{d11}
\end{equation}
From Eqs.~\eqref{d9} and \eqref{d11}, we obtain 
\begin{equation}
{\hat B}_D(t) \propto t^{-n_s/(1+n_s)}. 
\end{equation}
Then from Eqs.~\eqref{d10}  and \eqref{d11} we obtain 
\begin{equation}
\lambda_D (t) \simeq \lambda_S (t) \propto t^{1/(1+n_s)}, \label{d13}
\end{equation}
and from Eqs.~\eqref{d9} and \eqref{d11} we obtain 
\begin{equation}
{\hat B}_S(t) \propto t^{-n_s/(1+n_s)}. 
\end{equation}
Thus the exponents of the scaling laws for both fields are the same 
in the nonhelical case
if the direct cascade scaling law is realized with the discussions in the above. 

\section{Derivation of the $\lambda_{\rm phys}^0/{\rm pc} \simeq B_{\rm phys}^0/(10^{-14} {\rm G})$ condition \label{appc}}

Here we show how the relation $\lambda_{\rm phys}^0/{\rm pc} \simeq B_{\rm phys}^0/(10^{-14} {\rm G})$ is derived. 
At the recombination, the eddy turnover scale for the visible magnetic fields is 
\begin{equation}
\lambda_{\rm et}^{\rm rec} \simeq \frac{v_A^{\rm rec}}{H^{\rm rec}} = \frac{B^{\rm rec}}{\sqrt{\rho^{\rm rec}_B}H^{\rm rec}} = \frac{\sqrt{3/8\pi} B^{\rm rec} m_{\rm Pl}}{\sqrt{\Omega_B/\Omega_{\rm DM}}\rho^{\rm rec}}  = \frac{30 \sqrt{3/8\pi} z_{\rm rec} B^{\rm rec} m_{\rm Pl}}{\pi^2 g_*^{\rm rec} \sqrt{\Omega_B/\Omega_{\rm DM}} (\Omega_{\rm DM}/\Omega_\gamma) T_{\rm rec}^4} ,
\end{equation}
where $\rho_B^{\rm rec}$ is the energy density of baryons at recombination,
$\Omega_B$,  $\Omega_{\rm DM}$, and $\Omega_\gamma$  are the density parameters of baryons, dark matter, and relativistic particles, 
respectively, $z_{\rm rec}$ is the redshift at recombination, and $g_*$ is the number of 
relativistic particles at recombination. 
We can evaluate that the coherence length of the magnetic fields are the comparable 
to the eddy turnover scale. 
Assuming that after the recombination magnetic fields evolve adiabatically~\cite{Durrer:2013pga}, $
\lambda_0 = z_{\rm rec} \lambda^{\rm rec} = z_{\rm rec} \lambda^{\rm rec}_{\rm et}, \quad B_0 = z_{\rm rec}^{-2} B^{\rm rec}$, 
we obtain the relation between the present magnetic field coherence length and strength as
\begin{equation}
\lambda_0 = z_{\rm rec}^4 \frac{30 \sqrt{3/8\pi} \Omega_\gamma B_0 m_{\rm pl}}{\pi^2 g_*^{\rm rec}\sqrt{\Omega_B \Omega_{\rm DM}}T_{\rm rec}^4} \sim 1 {\rm pc} \left(\frac{B_0}{10^{-14}{\rm G}}\right). 
\end{equation}
There are still discussions that turbulent plasma might appear again and 
the magnetic fields might still experience the cascade, which gives an uncertainty 
of the order of unity. 
However, taking into account the uncertainties coming from the amplitude of the velocity fields 
and the relation between the coherence length of magnetic fields and the eddy turnover scale, 
the above expression gives a good estimate
and consistent with the expressions in Refs.~\cite{Banerjee:2004df,Durrer:2013pga}.

\end{document}